\documentstyle[preprint,aps]{revtex}
\begin{document}

\title{Protein Foldability and Designability: General Physics and Pretty
       Chemistry}

\author{R. A. Broglia$^{1,2,3}$, G. Tiana$^{3}$ and H. E. Roman$^{1,2}$}

\address{$^1$Dipartimento di Fisica, Universit\`a di Milano,
         Via Celoria 16, I-20133 Milano, Italy.}

\address{$^2$INFN, Sezione di Milano, Via Celoria 16, I-20133 Milano, Italy.}

\address{$^3$The Niels Bohr Institute, University of Copenhagen,
         2100 Copenhagen, Denmark.}

\date{\today}

\maketitle

\smallskip
\noindent
Pacs numbers: 87.15.Da, 61.43.-j, 64.60.Cn, 64.60.Kw

\begin{abstract}
Making use of a simplified model for protein folding, it can be
shown that conformations which are particularly stable when
their energy is minimized with respect to amino acid sequence 
(in the sense that they display a large energy gap to the lowest
structrally dissimilar conformation), aside from leading to
fast folding, are highly designable (in the sense that many
sequences target onto it in the folding process). These results
are quite general, do not depend on the particular simmetries
displayed by the compact conformation chosen as native, and can
be obtained making use of a large class of contact energies.
On the other hand, the design of sequences which fold onto native
conformations displaying secondary structures and eventually 
tertiary simmetries, is a difficult task requiring a delicate
tuning of the contact energies.
\end{abstract}

\newpage 
\narrowtext 
A basic property of ``wild type'' proteins is that they fold on short call.
A second one is that they can accomodate numerous mutations that are neutral
with respect to structure changes.
Another is that they display, in the native conformation, secondary structures
and tertiary symmetries, features that define their biological specificity
\cite{creighton}.

Single domain notional proteins must display a large energy gap $\delta$
between the native state and the bulk of misfolded conformations that are
structurally dissimilar to the native state (calculated making use of a Monte
Carlo formalism), in keeping with the fact that fast folding is tantamount to a
value of the ``order parameter'' $\xi=\delta/{\bar U}\gg1$, where ${\bar U}$ is
the average value of the contact energies among the amino acids. In cases where
${\bar U}=0$, the order parameter to be used
is $\xi=\delta/\sigma$,  where $\sigma$ is the dispersion of the interaction
energies \cite{goldstein92,ssk94,sg92,s94,goldstein95,gas96,bryngelson,s97}.
The designed sequence which in the native conformation fulfills $\xi\gg1$, 
allows for billions of mutations leading to sequences which have the native
conformation as their non-degenerate ground state with energies lying inside
the gap and, more importantly, which fold to the native state in a time of the
order of that associated with the folding of the designed
sequence\cite{desig_1}. 

To design a sequence which folds fast and which is highly resilient to
point mutations is easily achived making use of a large class of
contact energies among the amino acids, and does not require the
native conformation to display "wild--type" secondary or tertiary
structures
\cite{s94,ueda,go,dill,sklonik,covell,godzik,socci,klimov,Shakh96,grosberg}. 

It is not difficult either to
derive contact energies 
which select, as the most designable conformations, those displaying 
``wild-type'' secondary structures \cite{wingreen}. What is very difficult, 
requiring a detailed
tuning of the contact energies, is to achive that the associated sequences
fold, let alone whether they do it fast or slow. 

To illustrate the above points we report, in what follows, results
obtained making use of a lattice model
\cite{goldstein92,ssk94,sg92,s94,goldstein95,gas96,bryngelson,s97}, 
where the differences 
between the amino acids are manifested in pairwise interaction energy of
variable magnitude and sign, depending on the identity of the interacting
amino acid. The configurational energy is
\begin{equation}
E = {1\over2}\sum_{i,j}^N U_{m(i),m'(j)}~
\Delta(\vert {\vec r}_i-{\vec r}_j\vert),
\label{hami}
\end{equation}
$\{ {\vec r}\}$ being the set of coordinates of all of the monomers describing
a chain conformation and $U_{m,m'}$ is the effective interaction energy 
between monomers $m$ and $m'$.
The quantity $\Delta(\vert {\vec r}_i-{\vec r}_j\vert)$
is a contact function. It is equal to one if sites $i$ and $j$ are at unit
distance (lattice neighbours) not connected by a covalent bond, and zero
otherwise. In addition, it is assumed that on-site repulsive forces prevent
two amino acids from occupying the same site simultaneously, so that
$\Delta(0)=\infty$.

We start considering the case of proteins displaying a large value of the
parameter $\xi$. A particular realization of this situation is provided by
a 20-letter amino acid chain, composed of 36 monomers and designed to
fold into the native conformation shown in Fig. 1a. The quantities
$U_{m,m'}$ used here correspond to the contact energies obtained from
a statistical analysis of real proteins data and were taken from Table
6 of ref.\cite{mj} ("disordered" M-J contact energies (MJ)$_d$). In this case, the average value of the contact energies
is ${\bar U}=0$, the corresponding standard deviation $\sigma\cong 0.3$.
The 36-mer chain denoted as S$_{36}$ and designed by minimizing, for fixed
amino acid composition, the energy of the native conformation with
respect to the amino acids sequence is shown in Fig.1b . In the units we 
use ($R T_{\rm room}=0.6$ kcal/mol), the energy of S$_{36}$ in its native
conformation is $E_{\rm nat}=-16.5$ and the value of the gap $\delta=2.5$,
yielding a value for the order parameter $\xi=8.3$.
Making use of Monte Carlo (MC) simulations it is found that at the temperature
$T=0.28$, optimal from the point of view of allowing for the accumulation
of statistically representative samples of the different simulations,
and at the same time leading to a consistent population of the native
conformation ($\approx 10\%$), the sequence S$_{36}$ folds into the native conformation
in $8\cdot 10^5$ MC steps.

The characterization of the role played by the different amino acids in the
folding process of S$_{36}$ have been carried out in ref.\cite{my} (cf.
also \cite{aggreg}), by
using mutations. It was found that the 36 sites of the native conformation
(see Fig.1a) can be classified as ``hot'' (red bead, numbered 6, 27, and 30),
``warm'' (yellow beads numbered 3, 5, 11, 14, 16 and 28), and ``cold'' (the
rest of the beads, coloured green) sites. On average, mutations on the 27
"cold" sites yield sequences that still fold to the native structure (neutral
mutations), although the folding time is somewhat longer than for
S$_{36}$. Sequences obtained from mutations on the six warm sites fold,
as a rule, to a unique conformation, sometimes different but in any case
very similar to the native one. Mutations on the three hot sites lead,
in general, to complete misfolding (denaturation) of the protein.
These results were found not to depend on the possible structures
displayed by the native conformations, nor on any three-dimensional
properties of the native conformation, aside from the general
requirement of being compact. In fact, making use of the (MJ)$_d$
contact energies,
any compact structure is an equally good native.
Similar results as those discussed above were obtained making use of a
random force with the same average and standard deviation of the
(MJ)$_d$
contact energies, as well as with the Go
model \cite{go}.

Because of the large value of $\delta$, and the many possible sites available
to introduce neutral mutations, there are of the order of $10^9-10^{10}$
sequences S$_{36}'$ obtained from S$_{36}$ through single- and multiple- amino
acid concentration conserving mutations (swapping of amino acids) which have an
energy, when calculated in the native conformation, lying within the gap
$\delta$. The sequences S$_{36}'$ fold into the native conformation shown in
Fig. 1a, in times of the order, although somewhat longer, than that associated
with the folding of S$_{36}$ \cite{desig_1}. These results indicate that the
native structure, which,  as mentioned above, can be any compact structure, is
highly designable.

A revealing example of the difficulties met in trying to design fast
folding, highly designable proteins
with "wild type"-like secondary structures is provided by 
the results of ref. \cite{wingreen} (cf. also \cite{kardar}),
where a complete enumeration and energy calculation of
all compact configurations of 2 letters chains containing 27 monomers
was presented. The designability of each compact structure was measured
by the number of sequences S$_{27}$ that can design the structure, that is,
sequences that posses the structure as their non-degenerate ground
state. It was found that highly designable conformations posses
``protein-like'' secondary structures and even tertiary symmetries,
and are thermodynamically more stable than other conformations.

In keeping with the fact that the basic test a notional protein should pass to
qualify as such is to fold to the native structure in a "short time" (typically
of the order of $10^5-10^6$ MC steps), we have calculated the dynamics of a
number of the sequences designed by Li {\it et al.} \cite{wingreen}. We have found
that neither the sequences associated with the poorly-designable (cf. Fig.2(c))
nor with the highly-designable (cf. Fig.2(a)) conformations fold.  Such
sequences are thus unlikely candidates for coding functional proteins. 

The reason for these results is to be found on the fact that all the structures
of the two letter chain display a gap $\delta$ which is negative. In fact
elongated, poorly designable and thus structurally dissimilar configurations
exist (cf. Fig.2(b)), sampled by the Monte Carlo simulations which, for a given
amino acid composition have an energy which is lower than that associated with
the original, highly designable configuration. Defining the energy gap $\delta$
as the energy difference  between the native and the lowest structural
dissimilar {\it compact} configuration, leads to a positive value of $\delta$
which is of the order of the average contact energies, and consequently to an
``order parameter'' $\xi$ of the order of 1.

At the ``microscopic'' level we have found that all the configurations of the
two letter 27mer chains display a very large number of "hot" sites. In fact,
essentially half of all sites are "hot" sites (cf. Fig.2). In other words,
introducing single point mutations in the most designable structures of
ref.\cite{wingreen} one finds that one, out of two,  leads to protein
denaturation (i.e., in the present case, makes unstable the native structure),
a behaviour not observed in ``wild-type'' single domain proteins. In fact, the
strategy of multi-domain proteins, where the folding of each domain is
controlled by a small number of strongly interacting amino acids, is used here
to keep in place secondary structures on a single domain protein.

A possible way to design a viable notional protein making use of the
3D, 27-monomer chains of ref. \cite{wingreen} 
could be to modify the contact energies used in this paper,
eventually introducing an angle dependence (cf. e.g. ref.\cite{lund1}) so as
to selectively change the value of $\delta/{\bar U}$. 
That is,
to force the ratio $(\delta/{\bar U})_{\rm non-desig}$ associated with the 
non-designable
structures to remain at the present value (of the order of unity),
while boosting the ratio $(\delta/{\bar U})_{\rm desig}$
 characterizing the designable
structures to become of the order of 10.

In keeping with this discussion we report results of MC simulations carried out
for the twenty letter 27mer making use of contact energies taken from Table 5
of ref.\cite{mj} ("ordered" Miyazawa-Jerningan contact  energies (MJ)$_o$).
Also of a U-matrix containing 210 matrix elements of the form
$U_{ij}=h_{ij}+\eta$, where $h_{ij}$ can be $-1$, $0$ or $1$ according to the
hydrophobicity or polarity of the ith and jth residues, while $\eta$ is a
random number taken from a Gaussian distribution centered around zero and with
standard deviation $0.34$ \cite{maksim}.
In these cases it is found that not all compact conformations can act as native
configurations. In fact, one has to carefully choose the ratio between local
and non-local contacts to obtain natives in which the designed sequence
displays a large value of $\xi$. Examples are shown in Fig.3, where the
corresponding conformations are associated with values of $\xi$ equal to $6.9$
(Fig.3(a)), $3.8$ (Fig.3(b)) and $0.2$ (Fig.3(c)).  Although these conformations
do not display secondary structures typical of "wild--type" proteins as
discussed in ref. \cite{wingreen}, they provide examples of the possible links
which, appropriately chosen contact energies may create between selected 3D
conformations and high entropy regions of the sequence space (large gap
$\delta$), that is between structure and foldability.

One can conclude that stability, designability and fast folding are basic {\it
physical} properties characterizing notional proteins, properties which depend
only on the large number
of degrees of freedom displayed by these system in sequence space,
while the presence of secondary
structures and tertiary symmetries are ``pretty'' expressions of detailed {\it
chemistry}.

\bigskip
\bigskip
\bigskip
Financial support by NATO under grant CRG 940231 is gratefully acknowledged. 
The help of the late Dr. N. D'Alessandro in modelling advice is gratefully
ackowledged.

\newpage
\narrowtext
\begin{figure}
\caption{ {\bf (a)} The conformation of the $36$--monomers chosen as 
the native state in the design procedure. Each amino acid residue
is represented as a bead occupying a lattice site. The design tends
to place the most strongly interacting amino acids in the interior of
the protein where they can form most contacts. The strongest
interactions are between groups D, E and K (compare to b), the last one
being buried deep in the protein (amino acid in site 27). {\bf (b)}
Designed amino acid sequence S$_{36}$. The color plots in figs. 1, 2
and 3 were obtained by using the graphic program of ref. \protect\cite{program}.}
\label{fig1}
\end{figure}
  
\bigskip
\begin{figure}
\caption{ 
  {\bf(a)} Most designable conformation of a $27$ monomer chain
  composed of two types of amino acids, namely hydrophobic (H) and
  polar (P) to which are associated the interaction energies
  $E_{HH}=-2.3$, $E_{HP}=-1.0$ and $E_{PP}=0$ \protect\cite{wingreen}.
  The corresponding
  sequence which minimizes this configuration with a fixed number of H and P
  residues (13 and 14 respectively) is
  $S_{1}\equiv$ HPPHPHPHPHPHPHPPHPHPHPPPHHH, and displays an energy gap
  $\delta=2.6$ to the lowest fully--compact structurally different
  configuration. Note however that non--compact conformations with energy
  lower than the energy of the native state have been found.
  Changes in the monomer sites plotted as
  red dots, correspond to the hot mutations discussed in the text
  Mutations in these sites increase the energy of the target structure by an
  amount
  of the order or greater than the so called energy gap $\delta$ which,
  in this case, is of the order of the standard deviation of the interaction
  energies
  ($\sigma=0.9$), {\bf(b)}
  Starting from a random, elongated configuration of the sequence denoted
  $S_1$, the chain compacts but does not fold, in the sense
  that it reaches different conformations which, within the statistic
  accumulated ($500$ coils followed through $10^8$ MC steps),
  it never targets into the most designable ("native") configuration
  shown in (a). Of all the random configurations targeted in the
  folding process by the sequence $S_1$, the one shown here is one,
  among many, which displays an energy which is lower than the most
  designable
  configuration shown in (a), {\bf(c)} Example of the so called
  'less-designable' conformation of Li  {\it et al.}
  \protect\cite{wingreen},
  The sequence $S_2\equiv$ PPHHHHPHPPHHPPHPHPPPHPPHPHH is found to
  minimize the energy of this conformation. Starting from
  random, elongated, configurations of
  this sequence the chain compacts but, again, never folded, in the
  sense explained in (b).}
\label{fig2}
\end{figure}

\begin{figure}
\caption{Examples of 27mers compact conformations. Sequence
energy has been minimized for each of them making use of a matrix
of the form $U_{ij}=h_{ij}+\eta$, where the Hermitian matrix $h_{ij}$ is
composed of four blocks of elements -1, 0, 0 and 1 respectively, while
$\eta$ is a random number taken from a Gaussian distribution
having average value zero and standard deviation $0.34$. The corresponding 
standard
deviation of the matrix $U_{ij}$ is then $0.65$. The designed
sequences S$_{27}$ displays, in the native conformation, values of
the order parameter $\xi=6.9$ (a), $\xi=3.8$ (b) and $\xi=0.2$ (c), 
respectively. In keeping with these results, configurations
(a) and (b) contain only $15\%$ and $7\%$ of "hot" sites (red beads)
respectively,
while almost $70\%$ of all sites are "hot" in configuration (c). 
While the sequences S$_{27}(\xi)$ associated with the
conformations (a) and (b) fold (in times of the order of $10^6$ MC steps),
the sequence associated with the conformation (c) does not. Consequently,
the compact conformation (c) does not qualify as a native conformation.}
\end{figure}

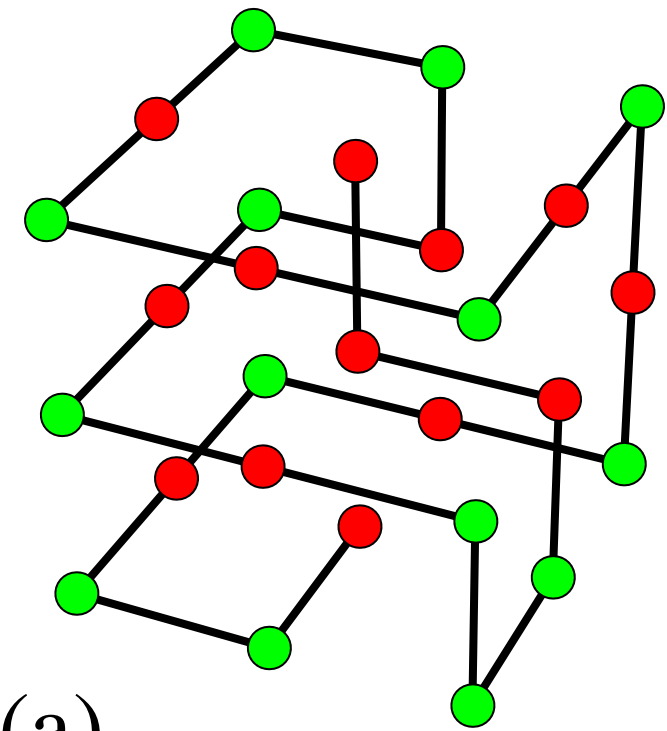
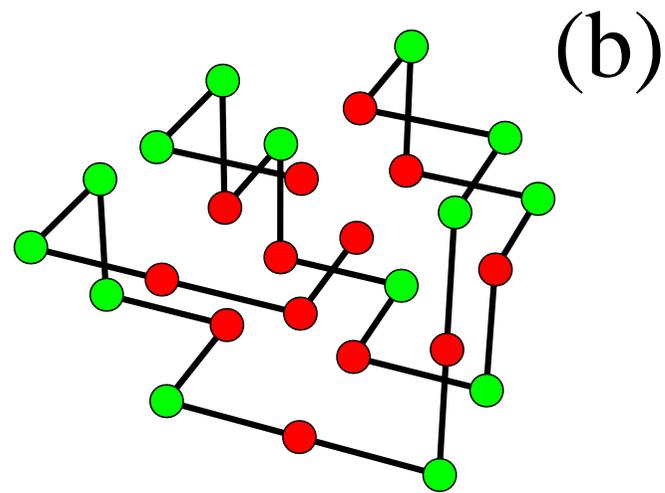
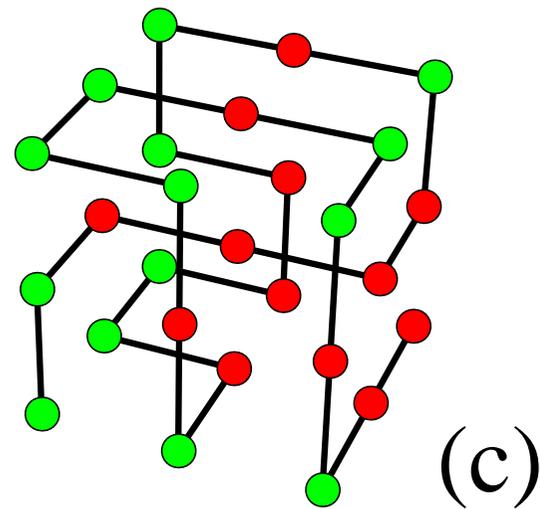
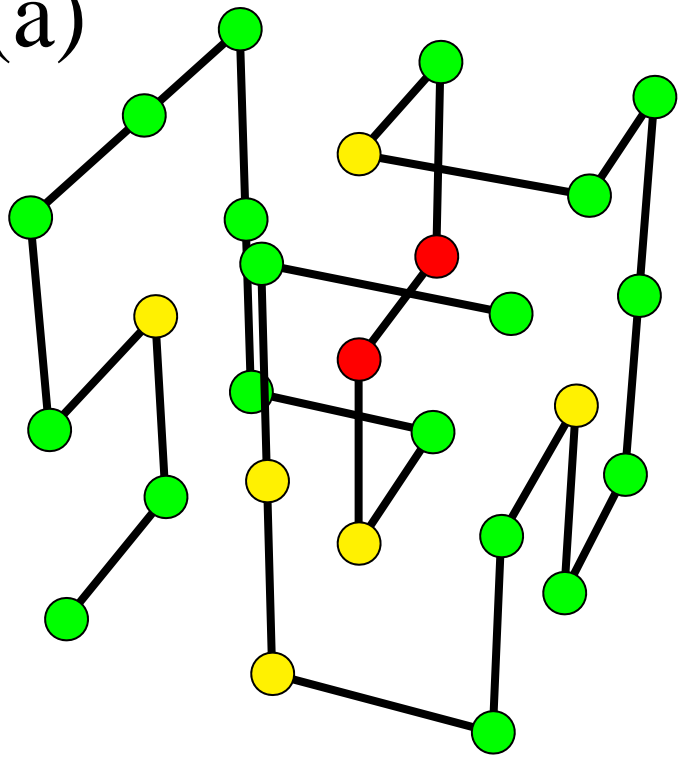
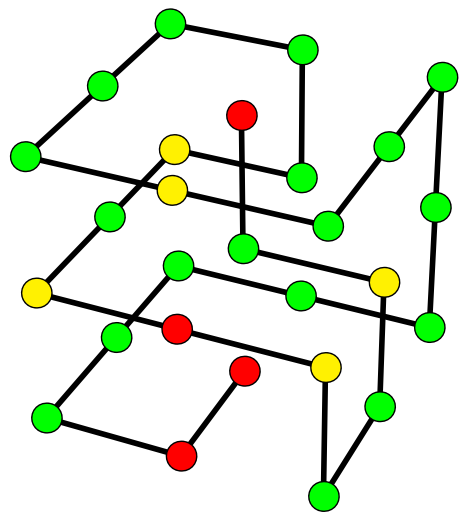
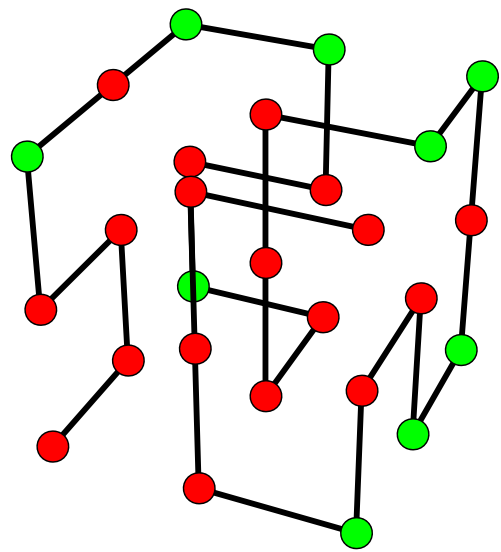
\end{document}